# Electron Microscopy Study of Core–Shell Nanowire Bending and Twisting

Spencer McDermott, Trevor R. Smith, and Ryan B. Lewis*

Department of Engineering Physics, McMaster University, L8S 4L7 Hamilton, Canada,
Email: rlewis@mcmaster.ca

**Abstract**

The spontaneous bending of core–shell nanowires through asymmetric shell deposition has implications for sensors, enabling both parallel fabrication and creating advantageous out-of-plane nanowire sensor geometries. This study investigates the impact of shell deposition geometry on the shell distribution and bending of GaAs–InP core–shell nanowires. Scanning and transmission electron microscopy methods are employed to quantify nanowire twisting and bending. A practical analytical electron tomography reconstruction technique is developed for characterizing the nanowire shell distribution, which utilizes the hexagonal nanowire shape to reconstruct two-dimensional cross-sections along the nanowire length. The study reveals that the orientation of the phosphorus beam with respect to the nanowire side facets induces significant variations in nanowire bending and twisting. The findings demonstrate the important role of crystallographic orientation during core–shell nanowire synthesis for engineering the shape of bent nanowire sensors.

**Introduction**

Bent nanowire heterostructures have garnered significant interest for sensors due to their scalable fabrication and the advantageous out-of-plane sensor geometry. To leverage bending for the mass production of nanowire devices, the fabrication processes must be well understood. Asymmetric core–shell nanowire heterostructures have been grown by physical vapor deposition processes—molecular beam epitaxy (MBE) [1–10] and electron beam evaporation [11–14]—as well as chemical vapor deposition with metal–organic MBE [15]. Nanowires are normally faceted to minimize surface energy [16–18], whereby both zinc-blende and wurtzite lattices can feature a hexagonal cross-section terminated at $\{1\bar{1}0\}$ and m-plane facets, respectively. Due to their non-cylindrical shape, equivalent directional deposition processes with fluxes incident on different nanowire crystallographic orientations can result in differing cross-sections and shell distributions [2,6,9,10]. The modeling of bending phenomena often assumes a cylindrical geometry [15,19,20], but the actual faceted geometries for crystalline nanowires have also been modeled [4,21–24]. Furthermore, nanowire twisting has been observed in nanowires on occasion [7,25,26], but the reason why some nanowires exhibit twisting while others to not is still unclear. Thus, further investigation is needed to determine the role of the cross-sectional orientation for directional nanoepitaxy of bent nanowires.

Cross-sectional transmission electron microscopy (TEM) characterization of nanowire heterostructures has been carried out on nanowire sections prepared by microtomy [9,27,28] and focused ion beam (FIB) milling [6,10,23,29,30], allowing for characterization of the core–shell interface, shell distribution, and asymmetry in the distribution. Atom probe tomography [31–33] offers high three-dimensional (3D) spatial and compositional resolution; however, it has constraints concerning the size and shape of nanowires that can be examined—one of which is the requirement for straight nanowires. Electron tomography utilizing scanning transmission electron microscopy (STEM) with energy-dispersive X-ray spectroscopy (EDS) or electron energy loss spectroscopy (EELS) in combination with a specialized tomography holder allows for conventional tomography of heterostructures [34–36]. The necessity of the specialized holder arises due to the requirement for numerous (time-consuming) scans across a broad range of angles [37]. However, reconstruction techniques can help reduce the required angular scan range and the number of scans. For example, the knowledge of faceting geometry in crystalline nanomaterials has been employed to reconstruct 3D nanomaterials [38,39] and cross-sections along nanowires [7,40–43]. In bent nanowires, bending variations along the nanowire can occur due to variations in core or shell geometry [3,7,25,26,44]. To explain nanowires with twisting or bending, it is crucial to understand the local cross-section along the nanowire.

In this paper, we investigate the influence of shell deposition orientation on the bending behavior and shell distribution of asymmetric core–shell nanowires. GaAs–InP core–shell nanowires are synthesized using phosphorus-controlled nanoepitaxy with gas-source MBE [9]. Scanning electron microscopy (SEM) is employed to examine the nanowires' shape, bending, and side facet orientation. The orientation of the phosphorus beam with respect to the nanowire side facets induces variations in the nanowire shape, bending, and twisting. A practical analytical transmission electron tomography reconstruction technique is presented to characterize the nanowire shell distribution, which employs the hexagonal fateing to reconstruct the 3D profile. This method circumvents the 'missing wedge problem' [37]. requiring only two scans at moderatly-separated tilt angles, resulting in an efficient approach to nanowire tomography and reducing beam damage. This technique is used to study the variations in core and shell along the length of bent nanowires. Two-dimensional (2D) cross-sectional distributions are generated from the STEM data and modeled with linear elastic theory. Nanowire twisting is observed in some nanowires and explained by the minimization of strain energy with respect to the shell distribution, which favors two shell orientations, symmetric about $<1\bar{1}0>$ and $<11\bar{2}>$, respectively.

**Methods**

GaAs–InP core–shell nanowires were grown on patterned Si (111) substrates by gas-source MBE. The substrates were prepared by first depositing 30 nm of $SiO_2$ via atomic layer deposition (ALD) with a FlexAL ALD cluster system. Next, a hole pattern in the oxide was created using electron beam lithography (EBL), where AR-P 6200.3:Anisole (1:1) e-beam resist was spun at 6000 rpm for 1 minute. The resist was soft baked at 150°C for 1 minute before undergoing EBL with a Raith EBPG 5000+ system. The pattern was then developed with ZED-N50, and reactive ion etching was used to transfer the pattern into the oxide. Immediately before loading the substrates into a SVTA-MBE35 MBE system, the substrates were immersed in a solution of 1 part Fujifilm Buffered Oxide Etchant 10:1 (NH4F:HF with Fujifilm surfactant) and 9 parts water for 28 seconds to remove native oxide from the hole pattern. GaAs–InP core–shell nanowire heterostructures were subsequently grown by MBE with solid-source effusion cells providing the gallium and indium. The V-sources, $As_2$ and $P_2$, were introduced via Arsine and Phosphine gases thermally cracked at 1000 °C.

GaAs nanowire cores were synthesized via vapor-liquid-solid growth at a substrate temperature of 630 °C. A pre-deposition of Ga for 250 seconds was performed before initiating nanowire growth via the addition of an $As_2$ flux at a V:III ratio of 2. The GaAs planar growth rate was 0.125 µm/h. After 20 minutes of GaAs growth, the $As_2$ flux was increased to a V:III ratio of 5.6 and maintained for 1 hour, after which, the Ga flux was terminated and the conditions held for one minute to consume the Ga droplets before cooling and reducing the $As_2$ flux. The same core growth process was used for all samples. InP nanowire shells were subsequently deposited at a substrate temperature of 410 °C in the absence of substrate rotation, with the substrate oriented so that the $P_2$ source was incident at various angles relative to the nanowire facets. Shell growth was initiated by opening the In shutter at a planar InP growth rate of 0.25 μm/h with a V:III ratio of 10 for a duration of 130 seconds. The $P_2$ and In sources were separated by an azimuthal angle of 108°. The In shutter was closed to terminate the shell growth.

As-grown nanowire samples were examined in an FEI Magellan 400 SEM operating at an accelerating voltage of 1 kV and a current of 13 pA. The secondary electron signal was detected using an in-lens detector in immersion mode. The nanowires were imaged normal to the substrate and at a 30° tilt. Top-view images (normal to the substrate) were collected under a stage bias of 500 V. Transmission electron microscopy samples were prepared by ultrasonicating as-grown samples in isopropyl alcohol for 2 minutes before drop casting on lacy carbon TEM grids. These samples were measured in a Talos F200X TEM with a X-FEG source equipped with a double tilt holder. Imaging was done in high-angle annular dark field (HAADF) STEM mode. EDS line scans were collected perpendicular to the nanowire's axial direction using four Super-X SDD in-column detectors. At each chosen location along the nanowire, two EDS scans were conducted, rotated by 23±4° about the nanowire axis. Velox by Thermo Fisher Scientific was used to analyze the net intensity of the K-shell for Ga and the L-shell for In in the nanowire heterostructure. Linear elastic modeling was conducted using MathWorks MATLAB to model curvature and strain energy for various core–shell distributions from different $P_2$ beam orientations.

**Results and Discussion**

Figure 1 shows SEM images of GaAs–InP core–shell nanowires with shells grown under different substrate azimuthal orientations relative to the In and $P_2$ sources. The nanowires show a significant difference in the amount of bending for different $P_2$ beam alignments. Nanowires grown with the $P_2$ beam incident along $<1\bar{1}0>$ ($P_2$ beam alignment of 0°, c.f. Figure 1a) exhibit greater overall bending compared to samples grown with 10°, 20°, 30° deposition angles (c.f. Figure 1b-d, respectively). For all nanowires observed on the samples, the bending direction is approximately in line with the incident $P_2$ beam direction, as previously reported [9]. The average curvature of the nanowires is plotted in Figure 1e, showing a monotonic decrease by about a factor of two as the deposition transitions from on the facet to on the edge between facets—from 0° to 30°. This illustrates a dramatic role of the flux orientation, not just for dictating the bending direction but also the amount of bending.

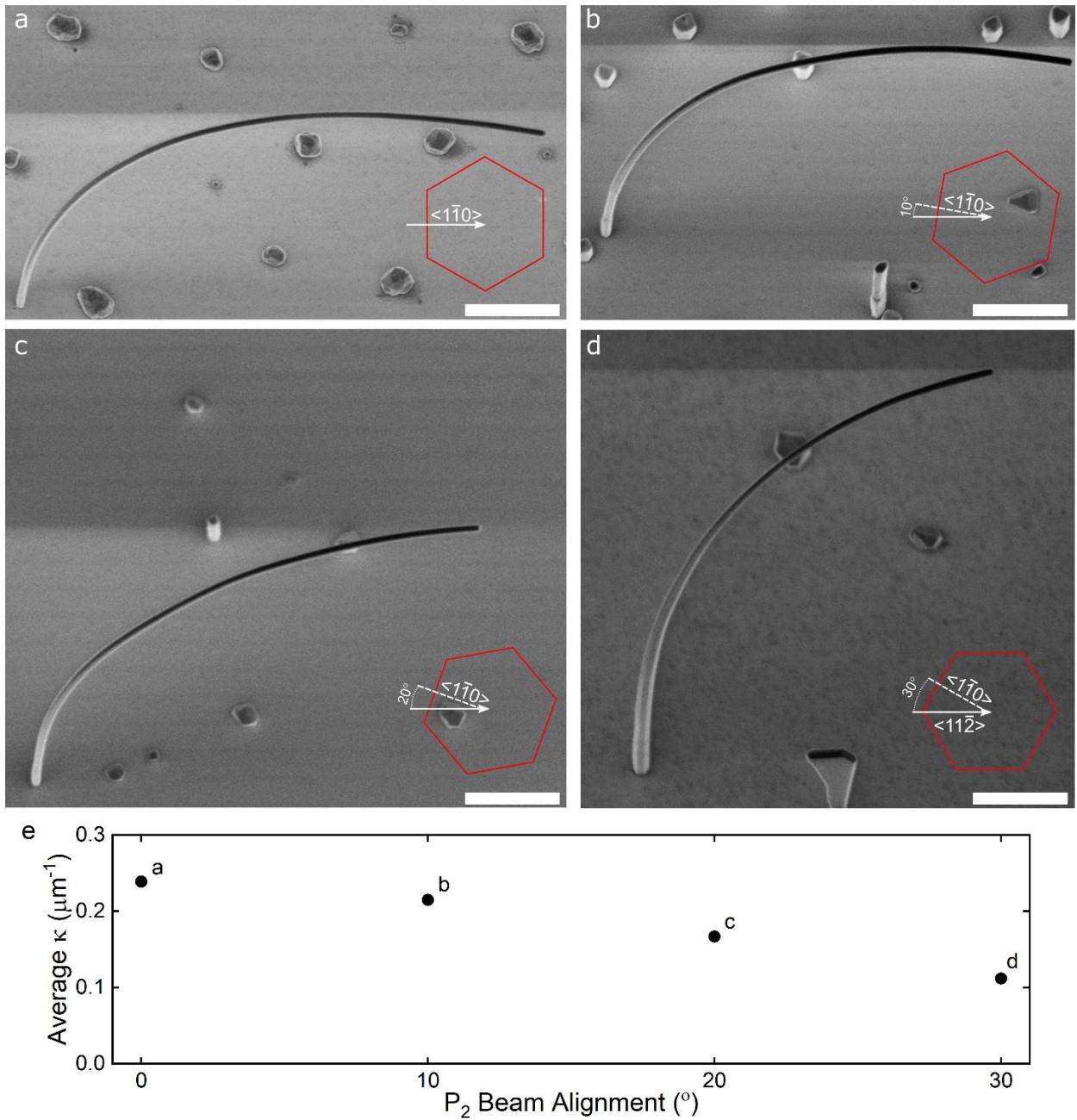

Figure 1. SEM images of core–shell nanowires with InP shells grown under various incident flux azimuthal angles. (a) Projected P$_2$ beam along the $<1\bar{1}0>$ direction, (b–d) P$_2$ beam offset from $<1\bar{1}0>$ by 10°, 20° and 30°, respectively. Images are taken at a tilt of 30° from the substrate normal and the insets show the angle of the incident P$_2$ beam with respect to the nanowire shape. All scale bars are 1 µm. (e) Average curvatures measured in (a-d) plotted with respect to P$_2$ beam alignment relative to $<1\bar{1}0>$.

Figure 2 presents top-view, high-magnification SEM images of the four samples shown in Fig. 1. Magnified views are shown of the base, mid-section and tip of each wire, where the dashed lines indicate the intersection of side facets. In Figure 2a—P$_2$ beam alignment along $<1\bar{1}0>$—no twisting is observed with side facets exhibiting a consistent orientation of $<1\bar{1}0>$ at the base, middle and tip of the nanowire. For the nanowire with P$_2$ beam offset by 10° (Figure 2b), the nanowire exhibits twisting. Specifically, at the base of the nanowire, the facet orientation is consistent with the expected bending/deposition direction, while the middle and tip of the nanowire indicate that the nanowire as rotated such that the $<11\bar{2}>$ (intersection of the facets) is now in the middle of the nanowire (pointing up). Similarly, for the nanowire grown with a 20° P$_2$ beam offset (Figure 1c), the facet orientation mid section and tip have $<11\bar{2}>$) pointing up, contrasting the geometry at the base of the nanowire and thus indicating that twisting has occurred. The nanowire grown with a 30° P$_2$ beam offset (Figure 1d) exhibits no twisting and is bent along the $<11\bar{2}>$ orientation from base to tip. For samples grown with 10° and 20° P$_2$ beam offsets, the nanowires twist such that the $<11\bar{2}>$ points upward—corresponding to twist angles of 20° and 10°, respectively. Only the nanowire grown without a P$_2$ beam offset shows bending along $<1\bar{1}0>$. Thus, the nanowires aligned at 0°and 30° do not twist and the nanowires with at 10°and 20° P$_2$ beam direction twist to orient the $<11\bar{2}>$ direction upward.

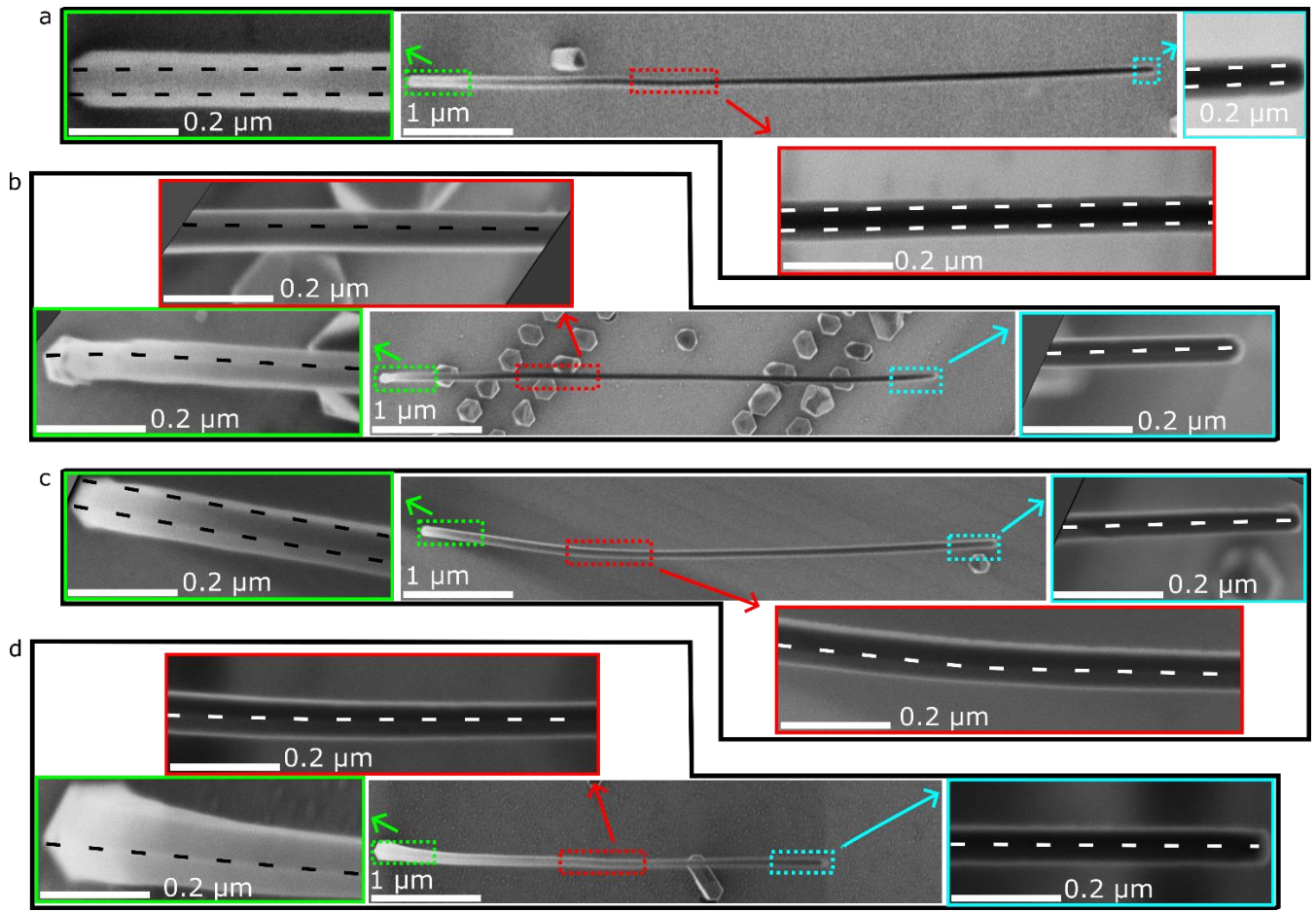

Figure 2. Top-view SEM Images of the bending series of Figure 1. Each sub-figure shows the nanowire in its entirety, along with three (5x) higher magnification images of the nanowire base (green), mid-section (red), and tip (cyan). The P$_2$ beam alignment in (a) is along the $<1\bar{1}0>$ direction and (b), (c), (d) are offset by 10°, 20°, and 30°, respectively. Black and white dashed lines denote the intersection of $\{1\bar{1}0\}$ side facets.

To quantify the core and shell distribution along the nanowire length, we developed a STEM tomographic method to reconstruct the core–shell geometry based on pairs of EDS lines scans collected at two different tilt angles about the nanowire axis. The STEM image in Figure 3a shows a nanowire grown with $P_2$ beam incident at 10° from $<1\bar{1}0>$, where EDS line scans are collected at ten segments along the length—shown as blue boxes. Each segment is measured at two angles, rotated by 23±4° around the nanowire axis. This is significantly less than the number of scans and the angular range required for traditional tomography [37], made possible by employing features known of nanowire geometry in the reconstruction—e.g., a hexagonal GaAs core, discussed below. Figures 3b and 3c show thickness profile reconstructions for a pair of scans at the same location along the nanowire. The EDS thickness profile reconstructions are created by assuming the faceting of the GaAs nanowire core—$\{1\bar{1}0\}$ facets terminating on the $<11\bar{2}>$ direction—is preserved for the InP shell, as seen in McDermott et al [9]. To convert the Ga line scans into thickness profiles, we assume hexagonal cross-sections for the GaAs cores and project the cross-section at different angles to best match the line scans. The intensity is scaled to align with each profile for each cross-sectional orientation and the orientation with the highest coefficient of determination is chosen as the correct profile of the nanowire core. The resulting profiles are plotted in Figures 3b and 3c and further detailed in the Supporting Information (SI). To fit the shell profile, it is assumed that both core and shell share the same faceting, so the slope of the rising edge of the core and shell thickness profiles are equivalent, as seen in Figure 3b–c. By fitting the In profile in this manner, the shell thickness profile is determined (see SI for further details).

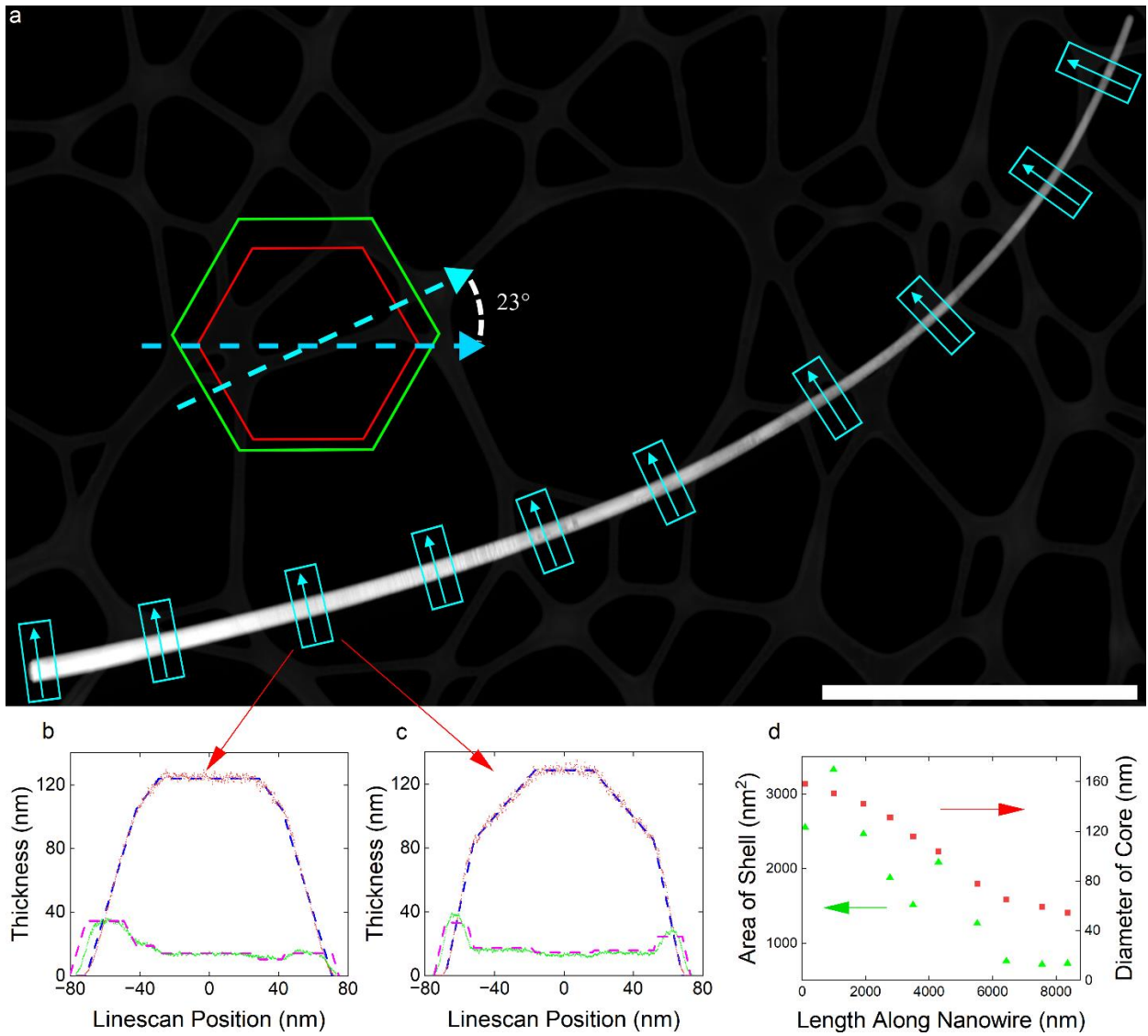

Figure 3. (a) Overview STEM image of a dispersed nanowire grown with incident $P_2$ flux 10° offset to $< 1\bar{1}0 >$. 10 EDS line scan pairs are displayed along the length, as indicated on the image by the blue arrow boxes. Pairs of scans are rotated by 23±4° about the nanowire axis, as illustrated in the top left corner of (a). The scale bar is 2 μm. (b–c) Exemplary thickness profiles extracted from a pair of EDS line scans. The blue and magenta dashed lines are the projection of the simulated core and shell, respectively. The EDS-measured core and shell data—scaled for appropriate thickness—are plotted in red (core) and green (shell). (d) Summary of measured diameter and shell area along the nanowire.

The 2D cross-sections are extracted from the EDS line profiles with three assumptions: 1) the shell is in contact with the nanowire core, 2) the compositions of the core and shell are pure GaAs and InP, respectivly, and 3) the shell has $< 1\bar{1}0 >$ terminating facets that intersect to enclose the shell. A single thickness profile will have multiple solutions that are unbounded along the direction from which the thickness profile was taken. A second thickness profile is thus required to extract a unique 2D cross-section with the above assumptions. To extract a profile, the unbounded shell profiles for each scan rotation are projected onto each other, where the best fit—projection with the highest coefficient of

determination—is chosen as the 2D profile (see SI for details). As shown in Figure 3b–c, there is excellent agreement between the core-shell modeling (dashed lines) and the EDS line scan data for both imaging directions. The core diameter and shell area along the nanowire is plotted in Figure 3d, indicating the nanowire core is highly tapered, with the core diameter decreasing from 158 nm at the base to 54 nm at the tip. On average, the shell area also decreases from the base to the tip, from 3300 nm² to 700 nm², respectively. The reduction in shell area is approximately proportional to the decrease in diameter, which is explained by the amount of In impinging per unit length being proportional to the nanowire diameter. It is noted that the maximum shell area is observed about 1 μm from the base and the decreasing area is not monotonic along the nanowire length.

The cross-sections reconstructed from the 10 EDS line scan pairs are displayed in Figures 4a–j. Averaging over all the segments, the two facets normal to 90° and 150° have relative shell thicknesses of 30.4±7.0% and 29.1±6.5% of the total thickness (summed of all facets), respectively. The opposite facets (normal to 270° and 330°) have relative shell thicknesses of 8.2±6.5% and 8.5±4.5%, respectively. The symmetry of this average cross section the P₂ direction was incident from 118.5° which is approximately aligned to the $<11\bar{2}>$ orientation. This means the nanowire has twisted 18.5° from its initial orientation.

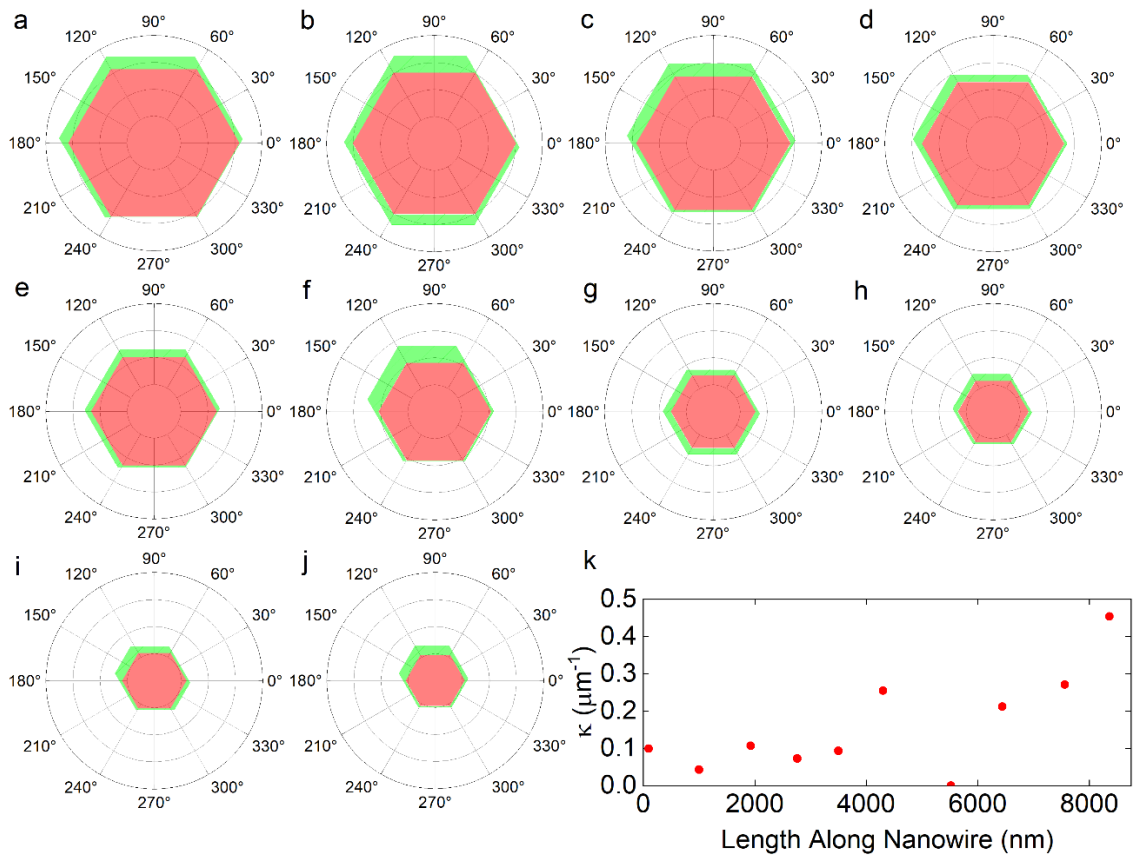

Figure 4. Extracted nanowire cross-sections from EDS measurements. The 10 cross-sections correspond to the 10 sections measured from the nanowire presented in Figure 3a, from base to tip (a–j). The red

area corresponds to the GaAs core and the green area to the InP shell. The radial line spacing is 25 nm. (k) Linear-elastic-theory-predicted curvature (κ) of the 10 cross-sections.

With the phosphorus beam incident from 120°, the two facets normal to 270° and 330° were shadowed from the P$_2$ beam and only receive P$_2$ flux from scattering—previously reported to be half of the direct beam flux [9,45,46]. Assuming the deposition to be proportional to the total incident P$_2$ flux, the shell thickness on facets normal to 270° and 330° suggest an average P$_2$ flux from scattering of 41–43% of direct beam. The calculated curvature is plotted for all ten cross-sections in Figure 4k using linear elastic theory with an assumed bending direction of $<11\bar{2}>$ from 120°. The calculated average curvature is 0.159 μm$^{-1}$, which is slightly higher than the value obtained from the STEM image of the same nanowire in Figure 3 (0.126 μm$^{-1}$), However, despite the agreement between the modeled and measured curvature for this nanowire, the average curvature is significantly lower than that observed in SEM for nanowires from this sample—the nanowire in Figure 1b has a curvature of 0.215 μm$^{-1}$. The cause of this discrepancy is unclear, and it is possible that the TEM-investigated nanowire is not representative of the average nanowire from this sample. Nevertheless, the fact that the modeled curvature agrees with the TEM-observed curvature on the same wire validates the EDS-extracted shell profile. The curvature locally modeled along the nanowire, as depicted in Figure 4k, reveals an approximately inverse relationship with core diameter (cf. Figure 3d). This inverse relationship has been previously demonstrated for nanowires with similar geometry in McDermott et al [3].

To explore the twisting phenomena and differences in bending, we simulated the bending of a 100-nm-diameter GaAs core with different InP shell configurations. For the model, the local shell thicknesses are assumed to be proportional to the P$_2$ flux received, based on are previous findings [9], and the scattered flux is assumed to be half of the direct beam flux and equally impinge on all side facets. Figure 5a displays the calculated curvature as a function of shell area for four P$_2$ deposition directions: the $<1\bar{1}0>$ direction, 10° offset, 20° offset, and the $<11\bar{2}>$ direction. The curvature can be categorized into three regimes: under-deposited, critically-deposited, and over-deposited, as detailed by McDermott et al [3]. Critical deposition corresponds to the peak curvature. As depicted in Figure 5a, the critical deposition does not significantly depend on the deposition orientation. For nanowires with P$_2$ flux along $<11\bar{2}>$, the critical deposition takes place at 5400 nm², which is marginally less than the 5900 nm² observed for nanowires oriented in the $<1\bar{1}0>$ direction. At critical deposition, the maximum curvature occurs when the phosphorus flux aligns with the $<1\bar{1}0>$ direction and diminishes as the nanowire orientation shifts to the $<11\bar{2}>$ direction, consistent with the experimental observations above. Importantly however, the difference in the modeled curvature is small, in sharp contrast with the experimental observation (Figure 1e). We speculate that the experimentally-observed reduction in bending might result from a variation in cross-sectional shell distribution, or favorable dislocation formation for certain shell geometries. Tomographic analysis reveals additional shell growth on side facets parallel to the phosphorus beam, which would reduce curvature. Further experiments are necessary to explore the origin of this discrepancy.

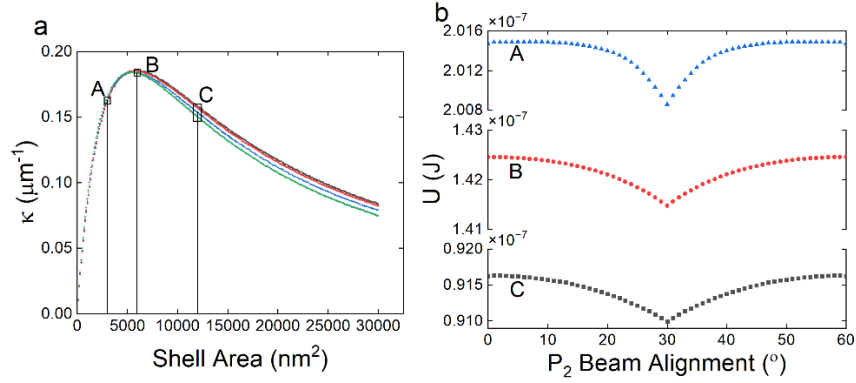

Figure 5. Linear elastic modeling of nanowire cross-sections with 100 nm diameter cores. (a) Curvature (κ) plotted as a function of shell area for four $P_2$ beam alignments offset from the $<1\bar{1}0>$ direction by 0° (black), 10° (red), 20° (blue), and 30° (green). Points A (3000 nm²), B (6000 nm²), and C (12000 nm²) show under-deposited, critically-deposited, and over-deposited shell areas, respectively. The strain energies for depositions corresponding to Point A (blue), Point B (red), and Point C (black) are plotted in (b) for $P_2$ beam alignments varying from 0° to 60°.

Figure 5b displays the strain energy plotted as a function of $P_2$ flux direction for under-deposited, critically-deposited, and over-deposited regimes (points A, B, and C in Figure 5a). For each curve, the minimum energy corresponds to the $P_2$ beam incident along $<11\bar{2}>$. Thus, the reason the nanowire twists to bend along the $<11\bar{2}>$ crystallographic direction involves energy minimization during shell deposition. Nanowires energetically favor growth in this direction and twist to reduce strain energy. Figure 5b shows two stable $P_2$ beam orientations: $<1\bar{1}0>$ and $<11\bar{2}>$. Twisting happens due to an energy gradient with the $P_2$ beam angle twisting towards an energy minima. The $<1\bar{1}0>$ $P_2$ beam alignment is a local minimum within ±1°, so most initial $P_2$ beam alignments will result in twisting towards the $<11\bar{2}>$ direction. Thus, the twisting in Figure 2 occurs because the nanowire orients itself in the $<11\bar{2}>$ direction to minimize strain energy. Interestingly, these findings are corroborated in Rossi et al [30] where InSb core–asymmetric shell geometry of As-poor and As-rich regions has a $<11\bar{2}>$ preferential bending orientation without any intentional beam alignment.

**Conclusion**

The twisting and bending behavior of nanowires subject to a directional deposition process at varying orientations relative to the nanowire facets have been explored. Nanowires aligned with $P_2$ flux (and shell growth) along the $<1\bar{1}0>$ direction exhibited nearly 2 times more bending compared to those oriented in the $<11\bar{2}>$ direction. Additionally, intermediate orientations—between $<1\bar{1}0>$ and $<11\bar{2}>$ directions—exhibit twisting toward the $<11\bar{2}>$ direction, which is shown to result from elastic strain energy minimization. A practical analytical electron tomography method is demonstrated, which allows for detailed examination of nanowire cross-sections, revealing local variations in diameter, shell area, and shell distribution. This technique can reconstruct nanowire cross-sections with only two scan directions utilizing known features from the nanowire geometry. This approach allows for practical imaging of bent nanowires. These findings demonstrate the importance of considering crystallographic orientation during bent nanowire synthesis, due to the significant impacts on bending and twisting.

**Acknowledgments**

We extend our thanks to the Canadian Centre for Electron Microscopy at McMaster University for providing electron microscopy facilities, the Quantum-Nano Fabrication and Characterization Facility at the University of Waterloo and the Toronto Nanofabrication Centre at the University of Toronto for providing facilities and technical support for substrate preparation, and McMaster's Centre for Emerging Device Technologies for providing MBE growth and fabrication facilities. We are grateful to Carmen Andrei for TEM support and Chris Butcher for SEM support. Financial support is acknowledged from the Natural Sciences and Engineering Research Council of Canada under Grant RGPIN-2020-05721.